\documentclass[]{aa}
\usepackage{epsfig}
\begin{document}
\def\gsim{\vcenter{\hbox{$>$}\offinterlineskip\hbox{$\sim$}}}
\def\lsim{\vcenter{\hbox{$<$}\offinterlineskip\hbox{$\sim$}}}
\title{The peculiar cluster HS 327 in the Large Magellanic Cloud: can OH/IR
       stars and carbon stars be twins?\thanks{based on observations obtained
       at the European Southern Observatory (La Silla, Chile)}}
\author{Jacco Th. van Loon\inst{1}, A.A. Zijlstra\inst{2}, L. Kaper\inst{3},
        G.F. Gilmore\inst{1}, C. Loup\inst{4}, J.A.D.L. Blommaert\inst{5}}
\institute{Institute of Astronomy, Madingley Road, Cambridge CB3 0HA, United
           Kingdom
      \and UMIST, P.O.Box 88, Manchester M60 1QD, United Kingdom
      \and Astronomical Institute, University of Amsterdam, Kruislaan 403,
           NL-1098 SJ Amsterdam, The Netherlands
      \and Institut d'Astrophysique de Paris, 98bis Boulevard Arago, F-75014
           Paris, France
      \and ISO Data Centre, Astrophysics Div., Science Dept.\ of ESA,
           Villafranca del Castillo, P.O.Box 50727, E-28080 Madrid, Spain
}
\offprints{Jacco van Loon, \email{jacco@ast.cam.ac.uk}}
\date{Received date; Accepted date}
\titlerunning{Can OH/IR stars and carbon stars be twins?}
\authorrunning{van Loon et al.}
\abstract{The obscured OH/IR star IRAS05298$-$6957 in the LMC was recently
noticed to be member of the small double cluster HS 327 that also contains a
carbon star (van Loon et al., 1998, A\&A 329, 169). Hence they are coeval and
have (nearly) the same progenitor mass, which can only be understood if Hot
Bottom Burning (HBB) has prevented IRAS05298$-$6957 from being a carbon
star.\\ We present extensive visual and near-IR photometric data for $>10^4$
stars in and around HS 327, and spectroscopic data for some of the brightest
AGB stars amongst these. Colour-magnitude diagrams are used to estimate the
age for the cluster and its members, and luminosities are derived for the
stars for which spectra have been obtained. The age for IRAS05298$-$6957 and
the carbon star is estimated to be $\sim200$ Myr. This corresponds to a
Main-Sequence progenitor mass $\sim4.0$ M$_\odot$ --- the first direct
measurement of the lower mass threshold for HBB. This agrees with stellar
evolution models that, however, fail to reproduce the low luminosity of the
carbon star.
}
\maketitle

\section{Introduction}

Stars of intermediate mass ($\sim1$ to 8 M$_\odot$) evolve along the
Asymptotic Giant Branch (AGB) before ending their life as a white dwarf (Iben
\& Renzini 1983). The energy production during the final part of the AGB
ascent normally takes place in a hydrogen shell surrounding the degenerated
carbon-oxygen core, but episodically an inner helium shell ignites (thermal
pulse = TP) which may cause material enriched with the products of nuclear
processing such as carbon and s-process elements to enter the convective
mantle. This is called $3^{\rm rd}$ dredge-up, and may result in a
photospheric C/O ratio exceeding unity, creating a carbon star. Thus the most
luminous AGB stars are expected to be carbon stars. These concepts seem to be
confirmed by the observation that in clusters in the Large Magellanic Cloud
(LMC) no M-type stars are found brighter than the brightest carbon star in
that cluster (Aaronson \& Mould 1985; Westerlund et al.\ 1991).

The relative efficiency of the $3^{\rm rd}$ dredge-up is higher for lower
metallicity, leading to the prediction of a large population of carbon stars
in the Magellanic Clouds (Iben 1981). Hence, the lack of observed carbon stars
with luminosities greater than $M_{\rm bol}=-6$ mag, a full magnitude below
the AGB-tip luminosity, came as a surprise (Iben 1981; Costa \& Frogel 1996).
Several reasons for this have been suggested, of which the most promising are
that carbon star formation is avoided by nuclear processing of carbon into
oxygen and nitrogen at the base of the convective mantle for the most massive
AGB stars (Hot Bottom Burning = HBB: Iben \& Renzini 1983; Wood et al.\ 1983),
or that luminous carbon stars become invisible at wavelengths shortward of
$\sim1 \mu$m due to obscuration by a circumstellar dust shell as a result of
intense mass loss on the TP-AGB. Recent theoretical work confirms the
occurrence of HBB in massive AGB stars with LMC metallicity (Boothroyd et al.\
1993; Frost et al.\ 1998; Marigo et al.\ 1998), and luminous obscured carbon
stars have recently been found in the LMC (van Loon et al.\ 1997, 1998,
1999a,b; Trams et al.\ 1999a).

The luminous OH/IR star IRAS05298$-$6957 (Wood et al.\ 1992) was noticed by
van Loon et al.\ (1998) to be situated in the core of the small cluster HS
327, and hence it may be possible to estimate its age by comparison of the
Colour-Magnitude Diagram (CMD) with theoretical isochrones. They also
serendipitously discovered a carbon star in the same cluster. This observation
suggests that carbon stars and (oxygen-rich) OH/IR stars may be coeval. More
TP-AGB stars have recently been identified in LMC clusters by Tanab\'{e} et
al.\ (1997).

We have obtained optical and near-IR imaging photometry for the HS 327 cluster
and surroundings, and (limited) spectroscopic observations for some of the
brighter stars, in order to derive an age for the stars in the HS 327 cluster.
The results are presented, and the implications for AGB evolution and carbon
star formation are discussed.

\section{Observations}

\subsection{Gunn gri-band imaging with the Dutch 0.9m}

The direct imaging camera at the Dutch 0.9m telescope at La Silla, Chile, was
used on the six nights of December 25 to 30, 1996, to obtain deep images of a
$3.77^\prime\times3.77^\prime$ region around the cluster HS 327, through Gunn
g ($\lambda_0=5148$ \AA, $\Delta\lambda=81$ \AA), r ($\lambda_0=6696$ \AA,
$\Delta\lambda=103$ \AA) and i ($\lambda_0=7972$ \AA, $\Delta\lambda=141$ \AA)
filters (Thuan \& Gunn 1976; Wade et al.\ 1979). The total integration time
amounts to $3^h45^m$ per filter, split into 5 minutes exposures to avoid
saturation and to allow for refocussing in order to reach the best image
quality. The pixels measure $0.442^{\prime\prime}\times0.442^{\prime\prime}$
on the sky, and stellar images on the combined (shift-added) frames have a
FWHM of $\sim1.5^{\prime\prime}$, though some individual frames show stellar
images of $\lsim1.1^{\prime\prime}$ FWHM. The CCD frames were reduced using
standard procedures within the ESO-MIDAS package. The photometry was
calibrated by means of regular observations of standard stars (see Appendix A
and Sect.\ 3.3 for more details).

\subsection{K$_{\rm s}$-band imaging with SOFI at the ESO/NTT}

SOFI at the ESO 3.5m NTT at La Silla, Chile, was used on February 15, 1999, to
obtain a K$_{\rm s}$-band ($\lambda_0=2.162 \mu$m, $\Delta\lambda=0.275 \mu$m)
image of HS 327 in order to identify the counterpart of IRAS05298$-$6957 and
to derive accurate bolometric luminosities for the brightest stars in the
vicinity of HS 327. Nine dithered images, each of $10\times2$ second
integration time were combined into one image with an effective integration
time of 3 minutes, at an airmass of 1.322. Standard near-IR observing and data
reduction techniques were employed. The sky background was derived from the
median of the individual frames, and subtracted. The pixels measure
$0.292^{\prime\prime}\times0.292^{\prime\prime}$ on the sky, and stellar
images on the final frame have a FWHM of $\sim1.0^{\prime\prime}$. For
photometric calibration the NICMOS standard star S 121-E ($K_{\rm
s}=11.781\pm0.005$ mag) was observed at an airmass of 1.224 by combining five
shifted images, each of $10\times2$ second integration time.

\subsection{I-band spectroscopy with EMMI at the ESO/NTT}

EMMI at the ESO 3.5m NTT at La Silla, Chile, was used on February 4, 1996, to
obtain low-resolution (R$\sim500$; grism \#4) spectra between $\sim0.61$ and
1.04 $\mu$m. The goal was to obtain a spectrum of IRAS05298$-$6957, but this
was unsuccessful. However, four stars in the vicinity of IRAS05298$-$6957 were
identified from their red colours on V and I-band acquisition images:
$(V-I)\sim3$ to 4 mag, compared to normal red giants that have $(V-I)\sim1$ to
2 mag. One of these is a carbon star, which spectrum was published and
discussed in van Loon et al.\ (1998). Here the results for the other three
stars are presented. All spectra were exposed for 5 minutes. The CCD frames
were corrected for the electronic offset (bias) and for the relative pixel
response (flatfield). Wavelength calibration was performed relative to He+Ar
lamp spectra. The sky-subtracted spectra were then corrected for the
wavelength dependence of the instrumental response as measured from a spectrum
of the standard star LTT 1020, and for atmospheric continuum extinction.

\subsection{I-band spectroscopy with EFOSC {\sc ii} at the ESO/3.6m}

EFOSC {\sc ii} at the ESO 3.6m telescope at La Silla, Chile, was used on July
5, 2000, to obtain a low-resolution (R$\sim400$; grism \#12) spectrum between
0.60 and 1.03 $\mu$m of a very red point source that was discovered on the
K$_{\rm s}$-band image, near HS 327. The spectrum was exposed 25 minutes. Data
reduction was analogous to the EMMI spectra, but the observed standard star
was LTT 7379. An appropriate filter was applied to remove charges from the
impact of energetic particles. Flexure was found to be negligible. The thin,
back-illuminated CCD \#40 in use with EFOSC {\sc ii} causes severe fringing at
wavelengths $\lambda\gsim0.8 \mu$m. The telluric emission lines of the sky at
dawn were especially bright and difficult to remove: preference was given to
an accurate stellar continuum determination for $\lambda\lsim0.7 \mu$m, which
in the crowded field allowed only small patches of sky near the object to be
taken.

\section{Analysis}

\subsection{Cluster morphology}

%
%
\begin{figure*}[tb]
\centerline{\hbox{
\psfig{figure=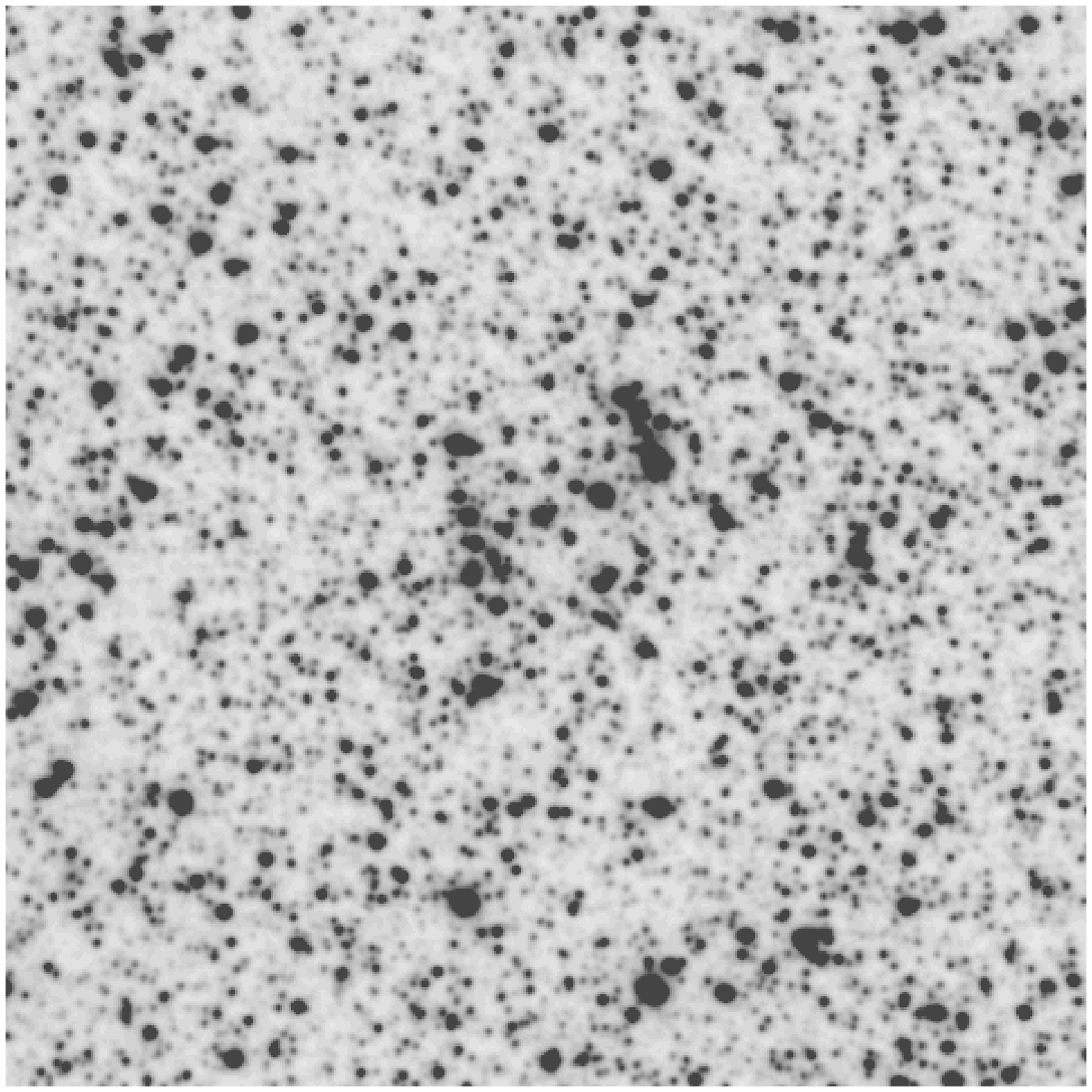,width=90mm,height=90mm,angle=-90}
\psfig{figure=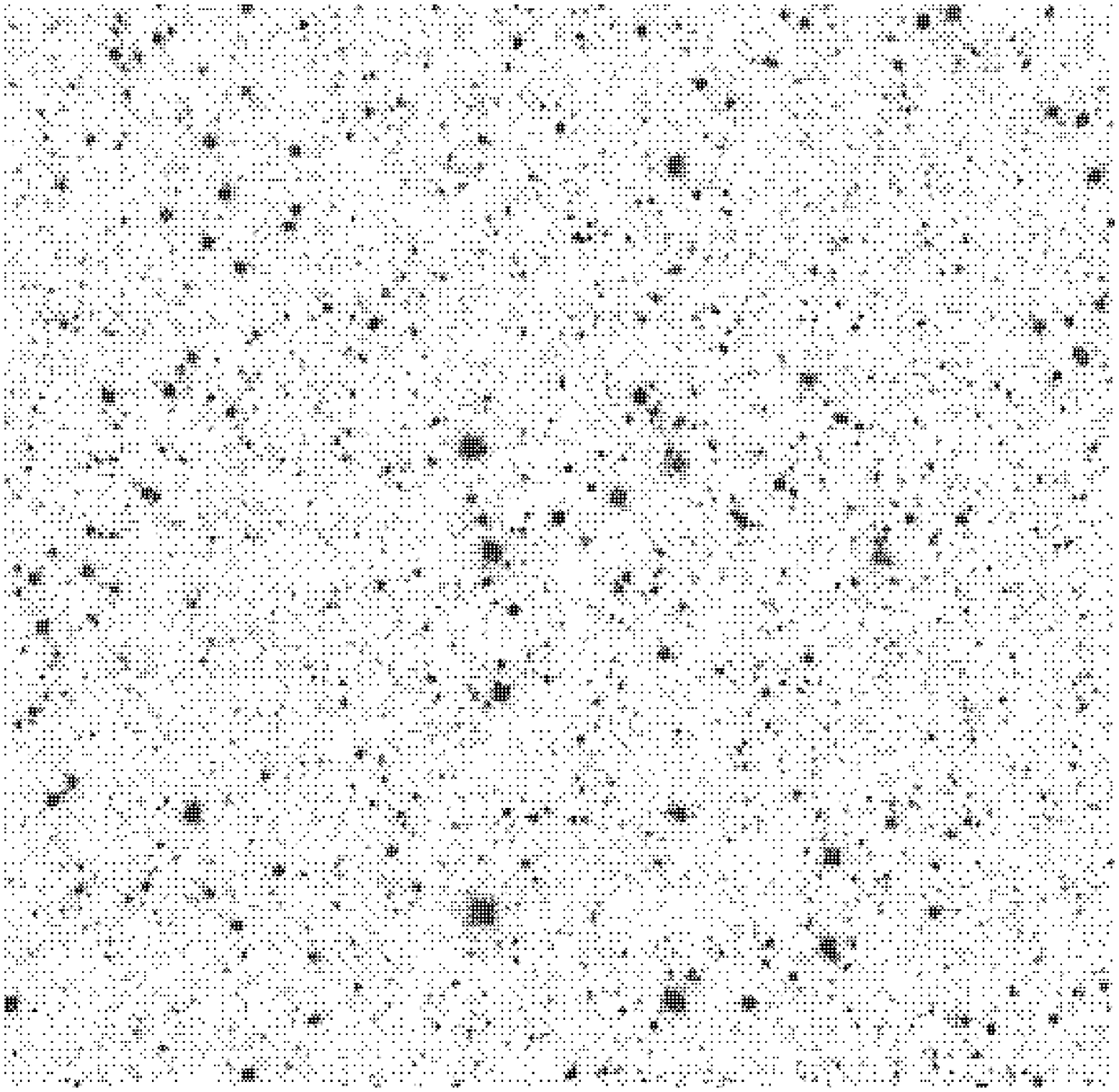,width=90mm,height=90mm,angle=-90}
}}
\caption[]{Dutch 0.9m Gunn-i (left) and NTT/SOFI K$_{\rm s}$-band (right)
images of the region around the LMC cluster HS 327. North is up and East to
the left; the edge measures $3^{\prime}40^{\prime\prime}$ on the sky.}
\end{figure*}

The open cluster HS 327 (Hodge \& Sexton 1966) is poor but rather compact,
measuring $\sim0.5^\prime$ across ($\sim7$ pc at the 50 kpc distance to the
LMC). Close inspection of the region revealed an accompanying loose cluster of
stars, that we call HS 327-E (East), at a distance of $\sim45^{\prime\prime}$
from the western cluster HS 327-W originally listed by Hodge \& Sexton (Figs.\
1 \& 2). The HS 327-E component may be identified with the small faint cluster
KMK 59 (Kontizas er al.\ 1988), but there is some confusion: SIMBAD's
(J2000.0) $\alpha=5^h29^m23^s \delta=-69^\circ55^\prime18^{\prime\prime}$
would indeed identify KMK 59 with HS 327-E, but Kontizas et al.\ themselves
give (converted to J2000.0) $\alpha=5^h28^m56^s
\delta=-69^\circ55^\prime28^{\prime\prime}$, placing KMK 59 $\sim2.5^\prime$
to the West of HS 327-E. The Second Generation Digitized Sky Survey does not
show anything obvious at Kontizas et al.'s position.

%
%
\begin{figure*}[tb]
\centerline{\hbox{
\psfig{figure=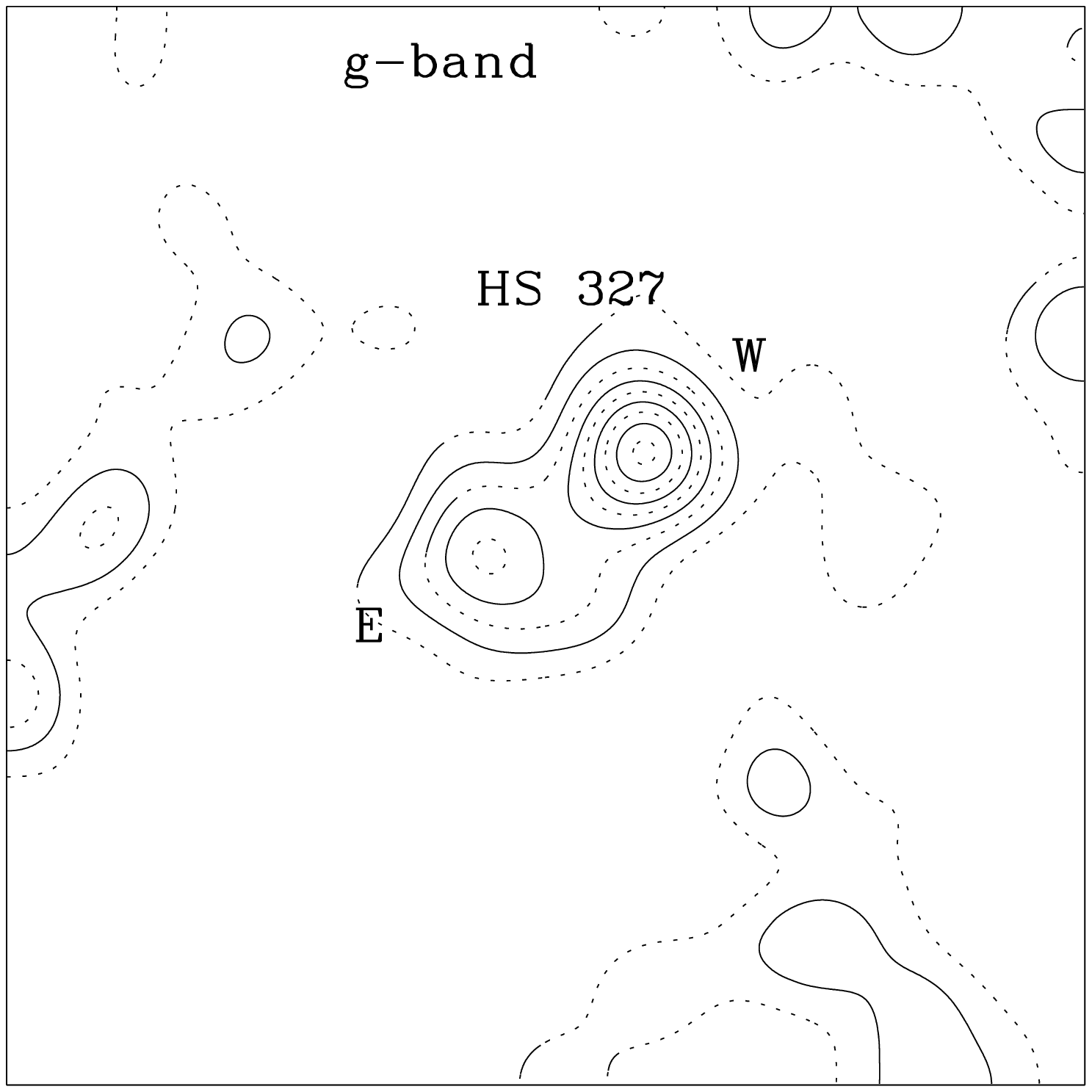,width=60mm,angle=180}
\psfig{figure=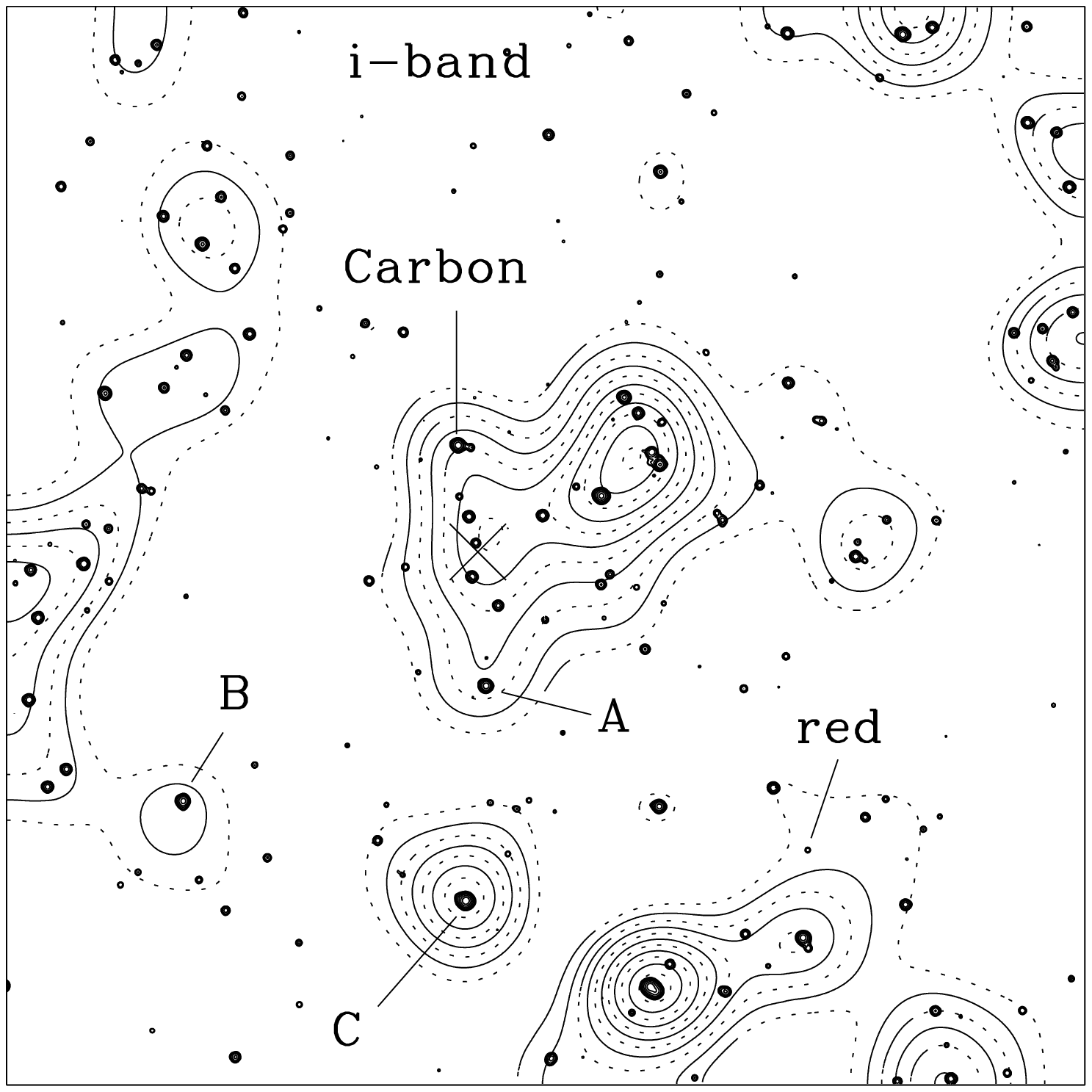,width=60mm,angle=180}
\psfig{figure=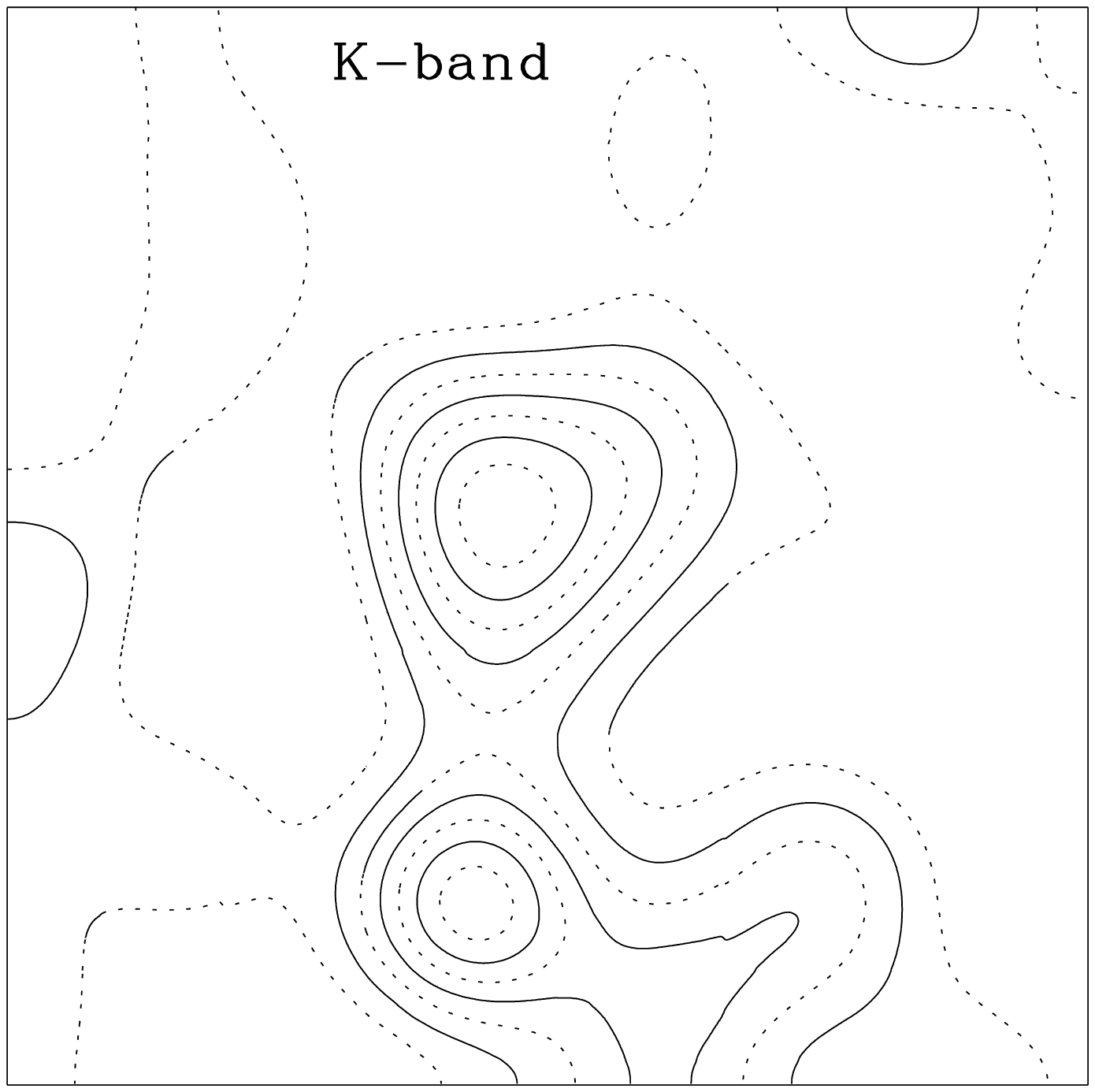,width=60mm,angle=180}
}}
\caption[]{Brightness contours in Gunn g (left), i (middle) and K$_{\rm s}$
(right). The i-band map is overlaid with the brightest stars, of which the
carbon star and the other three stars of which spectra were taken are
labelled, as well as the serendipitously discovered red star. The cross
indicates the position of IRAS05298$-$6957. Orientation and field size are the
same as in Fig.\ 1.}
\end{figure*}

The morphology of the cluster pair becomes apparent when brightness contours
are constructed from the images after smoothing by a Gau{\ss}ian filter with
$\sigma=17.5^{\prime\prime}$ (Fig.\ 2). HS 327 stands out most conspicuously
in the g-band because bright blue field stars are relatively rare. The visible
star density is highest in the i-band, where several individual bright (red)
stars add detailed structure to the morphology of the cluster as well as some
peaks not associated with the cluster. In the K$_{\rm s}$-band bright stars
become rare again, and only one of these (star ``C'') causes a peak rivalling
the cluster contours. What is most intriguing is that HS 327-W is blue, whilst
HS 327-E is red.

The brightest stars are overlaid on the i-band image (Fig.\ 2) to aid in
comparing the overall morphology with the location of individual sources. The
stars for which spectra have been obtained are labelled: the carbon star
(``carbon'': van Loon et al.\ 1998), three M-type stars (``A'', ``B'' and
``C''), and a very red carbon star (``red'') discovered on the K$_{\rm
s}$-band image, as well as the location of the OH/IR star IRAS05298$-$6957
(cross).

%
%
\begin{figure}[tb]
\centerline{\hbox{
\psfig{figure=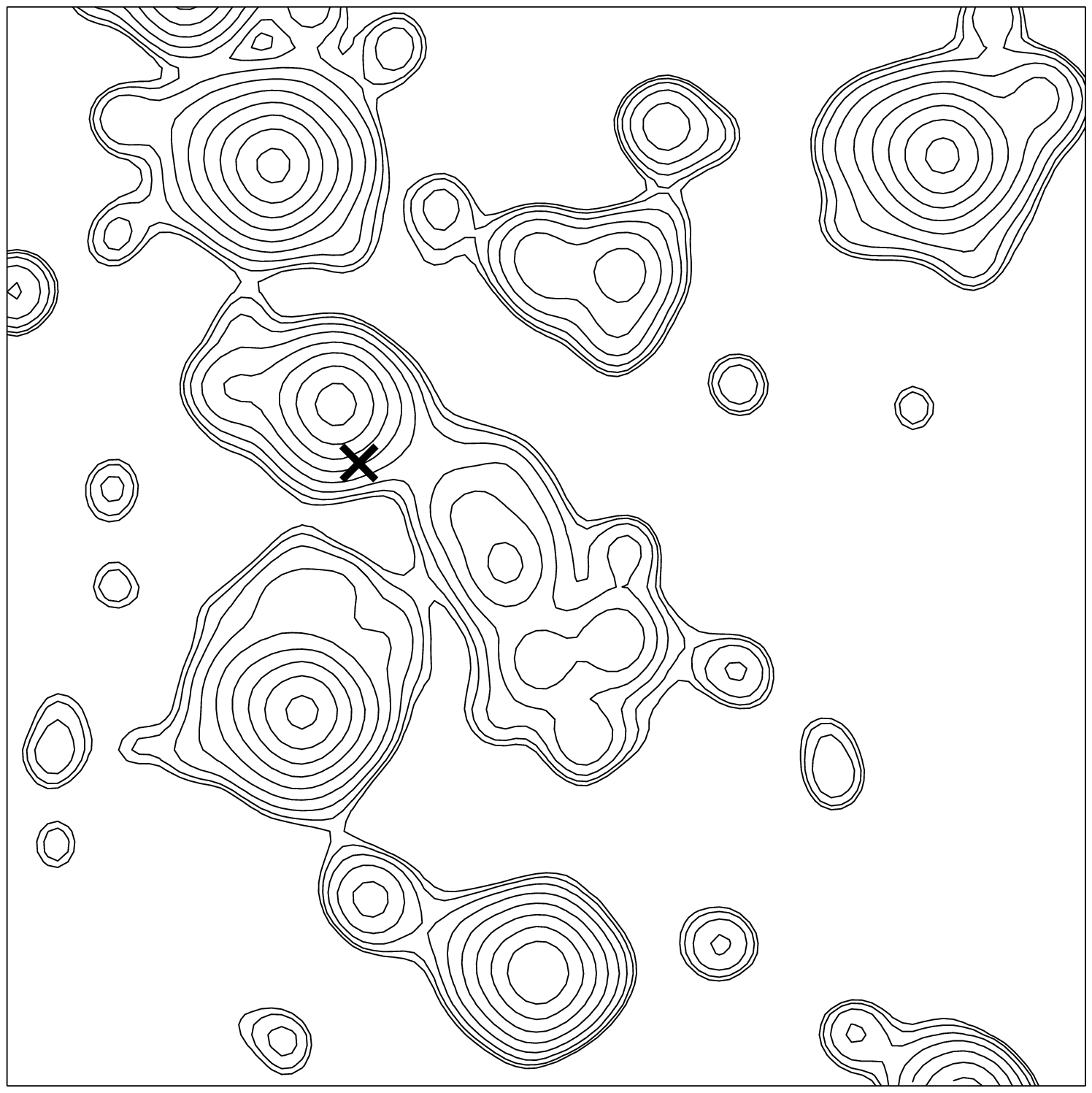,width=44mm,angle=180}
\psfig{figure=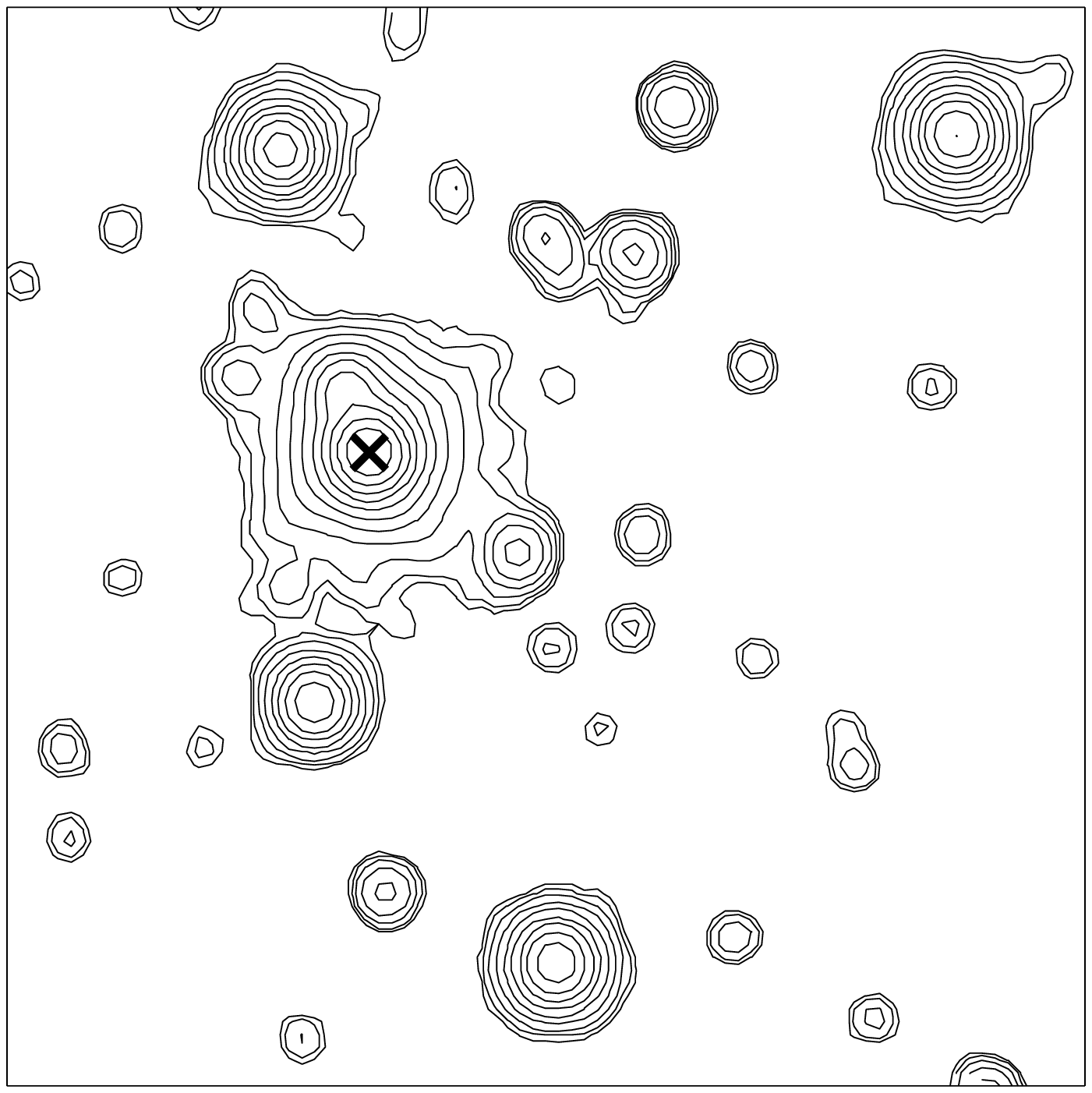,width=44mm,angle=180}
}}
\caption[]{Dutch 0.9m Gunn-i (left) and NTT/SOFI K$_{\rm s}$-band (right)
images of the immediate surroundings of IRAS05298$-$6957. North is up and East
to the left; the edge measures $24^{\prime\prime}$ on the sky. The cross marks
the position of the near-IR counterpart of IRAS05298$-$6957.}
\end{figure}

The near-IR counterpart of IRAS05298$-$6957 is very prominent in the K$_{\rm
s}$-band image, but (completely) invisible in the i-band image (Fig.\ 3).
Notice the blue star very near IRAS05298$-$6957. The 30 seconds i-band
acquisition image obtained with EFOSC {\sc ii} for the spectroscopy of the
very red star in the same field reached the same depth thanks to the greater
collecting area of the ESO/3.6m telescope and the $\sim1^{\prime\prime}$
stellar images.

\subsection{Cluster membership}

IRAS05298$-$6957, the carbon star and possibly also star ``A'' are seen in
projection against the Eastern part of the HS 327 cluster complex. How likely
is their physical association with the HS 327-E cluster?

Besides IRAS05298$-$6957 itself, there are five similarly bright mid-IR point
sources within a radius of $10^\prime$ from HS 327-E, and these are probably
not all obscured AGB stars. This corresponds to a surface density of $<70$
obscured AGB stars per square degree, or a $\lsim6$\% chance coincidence of an
obscured AGB star within $1^\prime$ of HS 327-E.

Similarly, there are nine known carbon stars including the cluster member, and
one additional possible carbon star, within a radius of $10^\prime$ from HS
327-E. This corresponds to a surface density of $\sim100$ carbon stars per
square degree, or a 10\% chance coincidence of a field carbon star within
$1^\prime$ of HS 327-E. The nearest known carbon star other than the cluster
member or the very red carbon star is SHV0530080$-$695949 at $2.7^\prime$.

Star ``A'' is a moderately luminous M-type star, which are less rare objects
than OH/IR stars or carbon stars. Indeed, the two similar stars ``B'' and
``C'' are found within $2^\prime$ from HS 327. With three such objects in a
field $>10\times$ the projected size of the cluster complex, the chance of
encountering star ``A'' within the cluster boundaries by coincidence is
$\lsim30$\%.

Thus it is highly likely that both the OH/IR star IRAS05298$-$6957 and the
cluster carbon star are physical members of HS 327-E, and probably the M-type
AGB star ``A'' too. The M-type stars ``B'' and ``C'' and the very red carbon
star (``red'') are considered not to belong to the cluster complex, as they
are all situated well outside the faintest brightness contour that isolates
the double cluster from the surrounding field.

\subsection{Spectral types}

%
%
\begin{figure}[tb]
\centerline{\psfig{figure=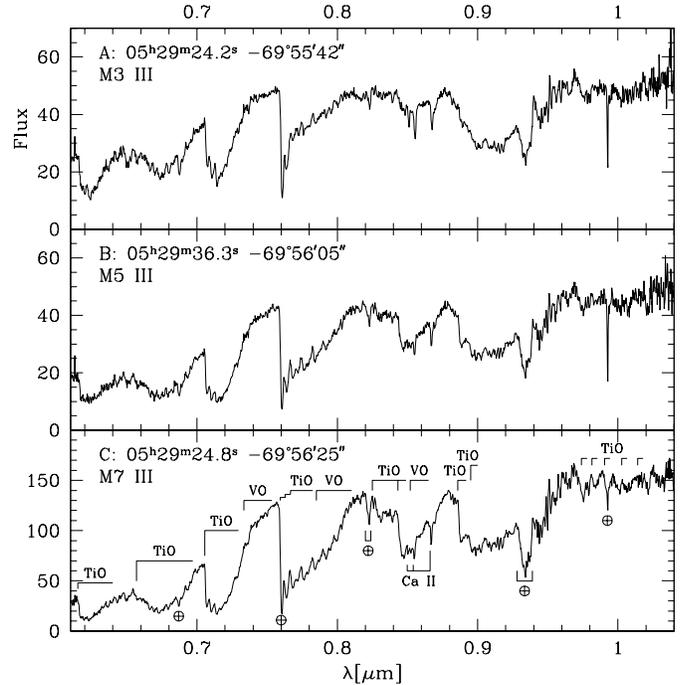,width=88mm}}
\caption[]{NTT/EMMI spectra of M-type stars near HS 327.}
\end{figure}

Spectra are now available for the brightest and reddest IR point sources in
the field around HS 327, which can be used to confirm that these are AGB stars
and to determine whether they are oxygen-rich (M-type) or carbon stars.

The spectra of stars ``A'', ``B'' and ``C'' are presented in Fig.\ 4 in
arbitrary flux units, together with their J2000.0 coordinates. The spectra are
classified by comparison with Turnshek et al.\ (1985). They are all M-type:
strong TiO absorption bands dominate, and absorption by VO becomes noticable
for the cooler star ``C''. The Ca {\sc ii} triplet is clearly visible in all
three stars, and reflects slight redshifts consistent with these stars being
members of the LMC. The Ca {\sc ii} strength and the estimated luminosities
(Section 3.8 and Table 2) indicate they are AGB stars.

The OH/IR star IRAS05298$-$6957 was already known to be an oxygen-rich AGB
star: spectra in the 3 to 14 $\mu$m region indicate the presence of
oxygen-rich circumstellar dust (van Loon et al.\ 1999a; Trams et al.\ 1999b),
and a 1.6 GHz spectrum of OH maser emission suggests LMC membership and a wind
velocity of $\sim11$ km s$^{-1}$.

%
%
\begin{figure}[tb]
\centerline{\psfig{figure=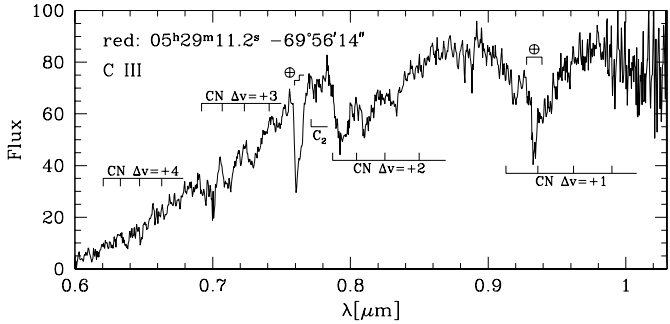,width=88mm}}
\caption[]{ESO 3.6m/EFOSC {\sc ii} spectrum of the very red carbon star near
HS 327.}
\end{figure}

The spectrum for the ``red'' star is presented in Fig.\ 5 in arbitrary flux
units, together with the J2000.0 coordinates. The reddened continuum shows
weak absorption of the strongest CN series and of C$_2$ near $0.77 \mu$m,
identifying this star unambiguously as a carbon star. The star may be only
moderately carbon-rich and/or rather warm. The absence of the Ca {\sc ii}
triplet around $0.86 \mu$m might indicate that the photosphere is of low
metallicity altogether. It is difficult to assign a spectral sub-type, but the
overall appearance and the lack of any conspicuous features that prove
otherwise suggest that it is of N-type, i.e.\ carbon enriched by $3^{\rm rd}$
dredge-up on the TP-AGB. (cf.\ Turnshek et al.\ 1985; Barnbaum et al.\ 1996).

For the spectrum of the cluster carbon star the reader is referred to van Loon
et al.\ (1998). They argue that it is an AGB star, which agrees with the
estimated luminosity (Section 3.8 and Table 2).

\subsection{Multi-object photometry}

Multi-object photometry was performed on the images using an implementation of
DAOPHOT (version {\sc ii}) and ALLSTAR (Stetson 1987) within ESO-MIDAS.

The Point Spread Function (PSF) was represented by an analytic Moffat function
with $\beta=2.5$ plus a quadratic lookup table to take into account variations
in the shape of the PSF across the frame. This PSF was established using
bright, relatively isolated stars in an iterative scheme involving subtraction
of contaminating neighbour stars. The multi-object photometry itself comprised
between several and a dozen passes (per filter) through DAOPHOT \& ALLSTAR,
each time subtracting already measured stars in order to also find and measure
fainter and/or blended stars. This resulted in the detection of
$\sim2.4\times10^4$ sources in each of the gri-bands, and $\sim1.0\times10^4$
stars in the K-band in the $3.7\times10^{-3}$ square degree area around HS327.

The photometric calibration was established by comparison of aperture
photometry on the standard star images with aperture photometry on a (few)
dozen bright stars in the images of the HS327 area (after subtracting
contaminating neighbour stars using PSF fitting). The multiple-object PSF
photometry was then scaled accordingly. The calibration of the gri-band
photometry, correcting for airmass dependencies of the photometric zero point,
is mainly based on the December 25 night but the photometric zero point has
been checked for the other nights too (see also Appendix A). Photometric
calibration errors are unlikely to exceed a few \%.

The completeness and photometric errors were estimated by a restricted number
of artificial star experiments. Incompleteness is mainly due to blending
and/or positional discrepancies as a result of crowding. The i-band
point-source extraction reaches a completeness level of 98\% at $i\sim18.75$
mag, dropping to 55\% and 5\% at $i\sim21.75$ and 23.75 mag, respectively. For
stars with $i\gsim22.25$ mag the error on the photometry is $\gsim1$ mag, with
magnitudes systematically too bright by $\gsim0.1$ mag. The g and r-bands go
$\sim0.75$ mag deeper. The K-band images reach $K_{\rm s}\sim19$ mag.

Point-sources were cross-correlated between the different filters after
geometric transformation (rotation and linear translation), using an iterative
scheme with a growing search radius, and rejecting extended sources on the
basis of the sharpness parameter returned by ALLSTAR.

\subsection{Colour-magnitude diagrams}

%
%
\begin{figure*}[tb]
\centerline{\vbox{
\psfig{figure=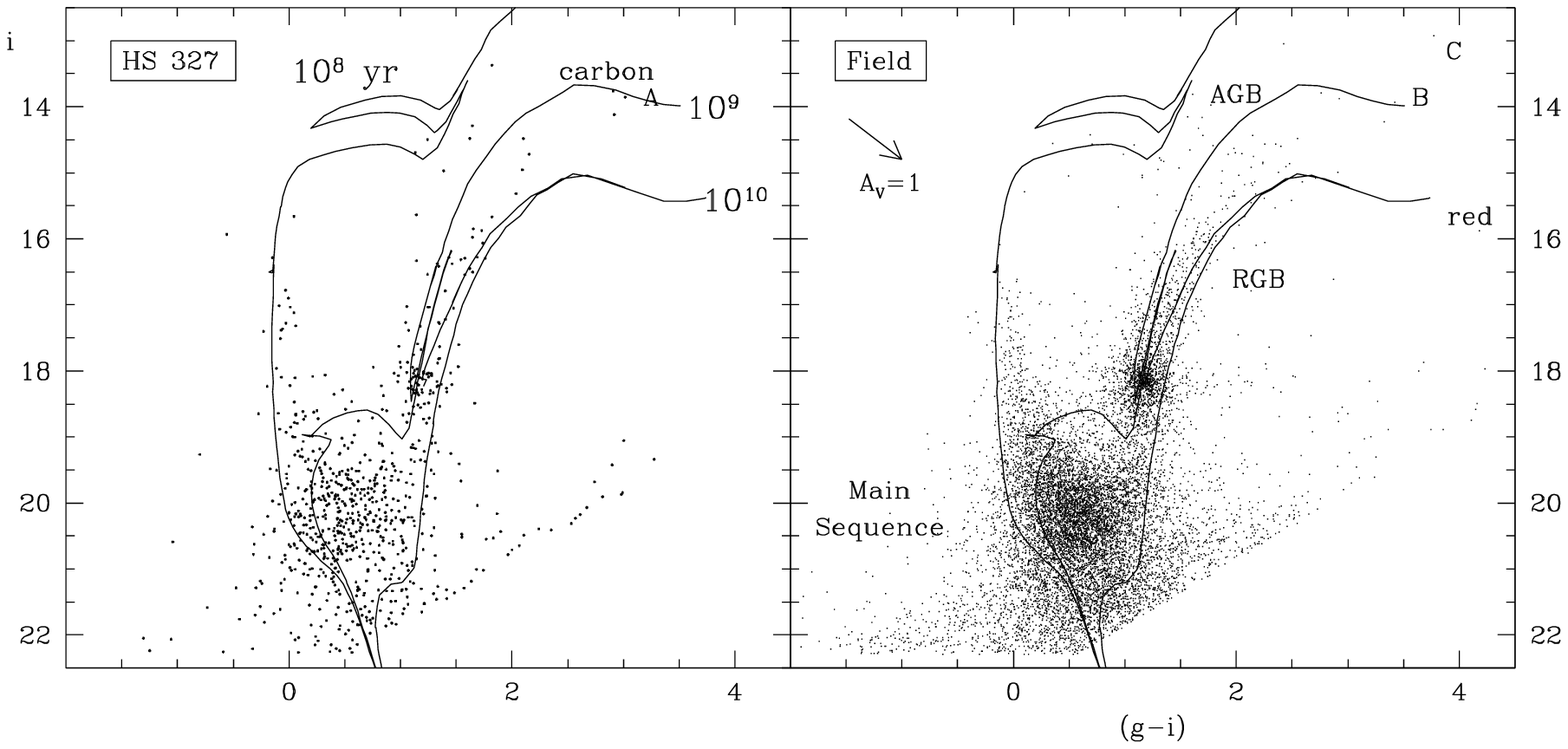,width=180mm}
\psfig{figure=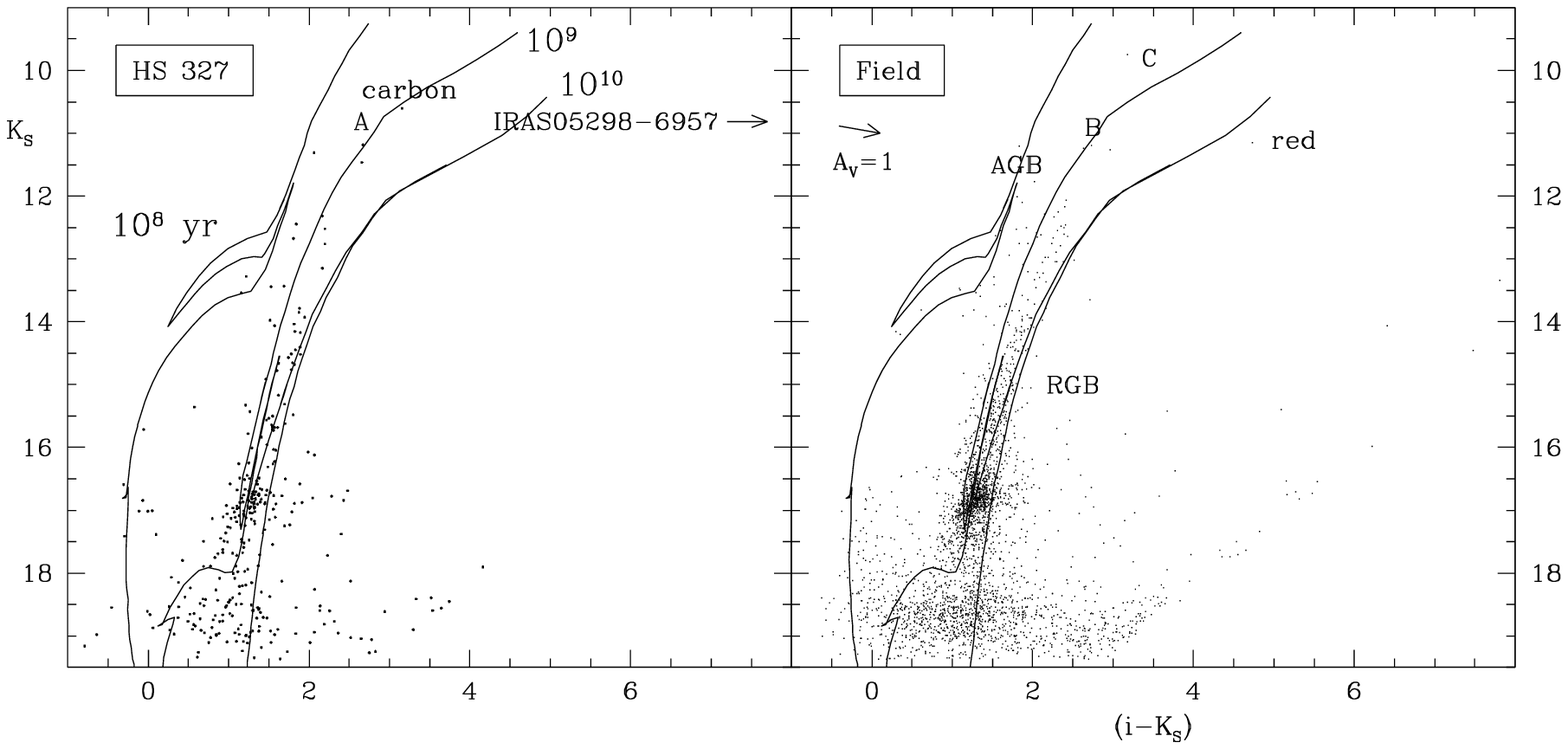,width=180mm}
}}
\caption[]{CMDs of $i$ versus $(g-i)$ (top) and $K_{\rm s}$ versus $(i-K_{\rm
s})$ (bottom) for the double cluster HS 327 (left) and surrounding field
(right). Stars whose spectra have been taken are labelled, and the location of
the AGB, RGB and Main Sequence are indicated. Isochrones (Bertelli et al.\
1994) are plotted for [metals/H]$=-0.4$, and ages of $10^8$, $10^9$ and
$10^{10}$ yr.}
\end{figure*}

The $i$ versus $(g-i)$ and $K_{\rm s}$ versus $(i-K_{\rm s})$ CMDs are given
in Fig.\ 6 for the HS327 cluster and surrounding field. The cluster boundaries
were defined by a rectangular box encompassing the faintest solid i-band
contour in Fig.\ 2. The red stars for which spectra have been obtained are
labelled, and the location of the AGB, RGB and Main Sequence are indicated.
The K$_{\rm s}$-band image barely reaches as faint as the Main Sequence. The
reddening vector is determined by convolving the extinction curve compiled by
Mathis (1990) with a Vega model (Kurucz 1993), yielding extinction
coefficients $A_\lambda/A_V=1.094$, 0.796, 0.621 and 0.112 for the gri- and
K$_{\rm s}$-bands, respectively. Stars of other spectral types have somewhat
different extinction coefficients, especially at shorter wavelengths. Kurucz
(1993) and Fluks et al.\ (1994) spectra are used to transform the isochrones
of Bertelli et al.\ (1994) into CMDs, taking into account the filter and CCD
response curves. The metallicity of the intermediate-age population is assumed
to be [metals/H]$=-0.4$ (typical for the LMC).

The field contains stars with ages ranging from $\sim0.1$ to $\sim10$ Gyr. The
age of the oldest stars is not very accurate because isochrones for stars of
such ages do not differ much in colours and the photometry becomes incomplete
above the Main Sequence turn-off for stars older than a few Gyr. Also, the
reddening of the stars with respect to the isochrones is a priori unknown,
although the general look of the CMDs suggests a visual extinction of no more
than a few 0.1 mag.

\subsection{Colour-colour diagrams}

Colour-colour diagrams have the potential to identify stars with particular
spectral signatures, such as carbon stars. In oxygen-rich cool stars the g-,
r- and i-bands include TiO absorption, with the i-band also including VO
absorption that is present in the coolest stars. In carbon stars the g-band
includes C$_2$ absorption, and the r- and i-bands include CN absorption
(stronger in the i-band). The K$_{\rm s}$-band mainly measures the continuum.

The carbon stars indeed exhibit somewhat different colours from the M-type
stars (Fig.\ 7). Carbon stars may have bluer $(g-i)$ colours than M-type stars
if they do not have very strong C$_2$ bands that absorb in the g-band, because
the g-band intensity of M-type stars is suppressed by TiO absorption. As can
clearly be seen in the synthetic, solar-metallicity M-type spectra from Fluks
et al.\ (1994), the coolest M-type stars have relatively red $(i-K_{\rm s})$
for their $(g-i)$ colours due to the TiO and VO absorption that affects the
i-band. The same happens for the carbon stars, but now it is due to the CN
absorption in the i-band.

There is only a small mismatch between the synthetic colours for M0 to M3
stars and the observed red giant sequence. This may be due to differences in
metallicity or slight inaccuracies in the synthetic models. The observed stars
redder than these early-M type stars --- including the three M-type stars A, B
and C --- are not represented by the synthetic sequence. This may be due to
the spectral-type dependence of the inter- and circumstellar reddening, which
is especially important for intrinsically red objects towards the blue
wavelengths. Large-amplitude pulsation of AGB stars may be responsible for
weaker TiO absorption bands in the i-band (Schultheis et al.\ 1998). It may be
noted that spectra are available for nearly all late-type AGB stars in the
field, and that there are no strong candidates for carbon stars that have
escaped notice other than the two identified.

%
%
\begin{figure}[tb]
\centerline{\psfig{figure=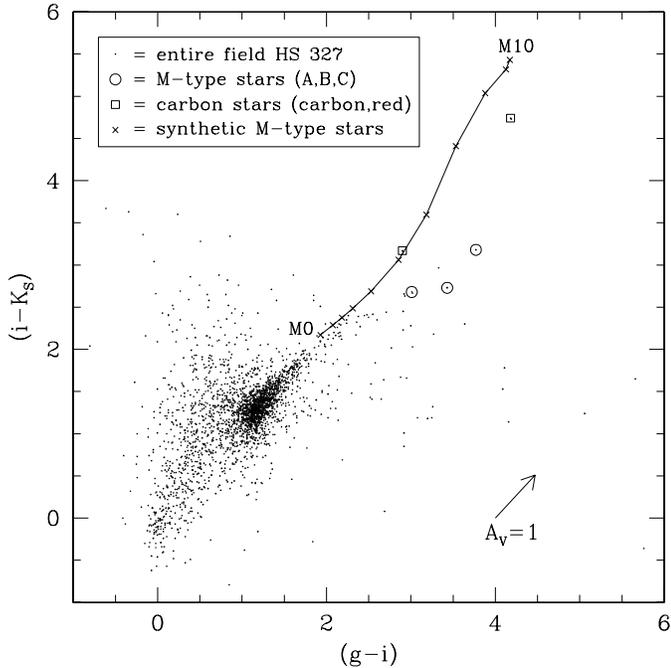,width=88mm}}
\caption[]{$(i-K_{\rm s})$ versus $(g-i)$ colour-colour diagram for the stars
in the field of HS 327. The M-type and carbon stars for which spectra have
been obtained are indicated by circles and squares, respectively. Synthetic
M-type spectra from Fluks et al.\ (1994) are plotted too (crosses, solid
line).}
\end{figure}

\subsection{Mid-IR photometry}

%
%
\begin{table*}
\caption[]{Optical, near- and mid-IR photometry for the six red stars in the
HS327 area. Coordinates have $\sim1^{\prime\prime}$ accuracy.}
\begin{tabular}{lccrrrrrrrr}
\hline\hline
Star               &
$\alpha$ (J2000.0) &
$\delta$ (J2000.0) &
g                  &
r                  &
i                  &
K$_{\rm s}$        &
[4.5]              &
[6.7]              &
[12]               &
[14.3]             \\
\hline
\multicolumn{11}{l}{\it Cluster members:} \\
IRAS05298$-$6957   &
$5^h29^m24.5^s$                       &
$-69^\circ55^\prime14^{\prime\prime}$ &
$\gsim22.00$       &
$\gsim22.00$       &
$\gsim22.75$       &
10.81              &
 7.47              &
 5.54              &
 4.76              &
 3.68              \\
carbon             &
$5^h29^m25.1^s$                       &
$-69^\circ54^\prime53^{\prime\prime}$ &
16.67              &
15.07              &
13.77              &
10.60              &
10.18              &
 9.94              &
 9.46              &
 8.74              \\
A                  &
$5^h29^m24.2^s$                       &
$-69^\circ55^\prime42^{\prime\prime}$ &
16.87              &
15.60              &
13.86              &
11.18              &
11.23              &
11.52              &
$\gsim11.40$       &
$\gsim10.80$       \\
\multicolumn{11}{l}{\it Field stars:} \\
red                &
$5^h29^m11.2^s$                       &
$-69^\circ56^\prime14^{\prime\prime}$ &
20.07              &
17.23              &
15.89              &
11.15              &
 9.18              &
 8.75              &
 8.55              &
 8.60              \\
B                  &
$5^h29^m36.3^s$                       &
$-69^\circ56^\prime05^{\prime\prime}$ &
17.36              &
15.83              &
13.93              &
11.20              &
11.10              &
10.80              &
10.77              &
11.18              \\
C                  &
$5^h29^m24.8^s$                       &
$-69^\circ56^\prime25^{\prime\prime}$ &
16.70              &
15.12              &
12.93              &
 9.75              &
 9.74              &
 9.60              &
 9.60              &
 9.71              \\
\hline
\end{tabular}
\end{table*}

Mid-IR images at wavelengths of 4.5, 6.7, 12 and 14.3 $\mu$ exist that cover
the HS 327 area. These were obtained with the CAM instrument onboard ISO for
the ISOGAL (Omont et al.\ 1999) and mini-survey (Loup et al.\ 1999) projects.
The details of observation and data reduction as well as the catalogue with
point source photometry will be described in a subsequent paper by Loup et al.

Only very few objects in the HS 327 area are bright enough in the mid-IR to
have been detected by these ISO observations. Amongst these, IRAS05298$-$6957,
the carbon star, the ``ABC'' stars and the very red star are all detected in
at least one of the mid-IR passbands, which can be used to better determine
their bolometric luminosities. These stars also constitute six of the seven
brightest stars in the K-band. Besides these stars that we had already
identified as bright red giants on the basis of shorter wavelength data, no
other such stars appear in the mid-IR images of the HS 327 area.

The mid-IR photometry is summarised in Table 1. Stars A and B and the carbon
star are too faint and/or too close to IRAS05298$-$6957 to have been included
in the Loup et al.\ catalogue that was compiled using a homogeneous automatic
multiple-object photometry procedure. Photometry is obtained by aperture
photometry on the original images. The magnitudes follow the IRAS/ISO
convention, with zero magnitudes of the LW1 (4.5 $\mu$m), LW2 (6.7 $\mu$m),
LW10 (12 $\mu$m) and LW3 (14.3 $\mu$m) filters corresponding to 181.8, 89.5,
34.7 and 20.7 Jy, respectively. Additional near- and mid-IR photometry and
spectroscopy for IRAS05298$-$6957 can be found in Trams et al.\ (1999). This
object was also detected by MSX (Price \& Witteborn 1995) at 0.27 Jy in Band
A (Egan et al.\ 1999), centred at 8.3 $\mu$m (range: 6.8 to 10.9 $\mu$m).
Inspection of the MSX images (http://www.ipac.caltech.edu/ipac/msx/msx.html)
reveals that IRAS05298$-$6957 was also detected in Band D, centred at 14.6
$\mu$m, and that the red field carbon star was marginally detected at
$\sim0.03$ Jy in Band A (below the nominal sensitivity limit). This is all
consistent with the ISO photometry. The optical, near- and mid-IR photometry
in Table 1 was gathered at different epochs, which is important because
IRAS05298$-$6957 --- and possibly the other red stars as well --- is a
long-period variable star with a K-band amplitude of $\sim2$ mag (Wood et al.\
1992). LW2 and LW10 observations are available for two epochs, for which the
photometry was averaged.

%
%
\begin{figure}[tb]
\centerline{\psfig{figure=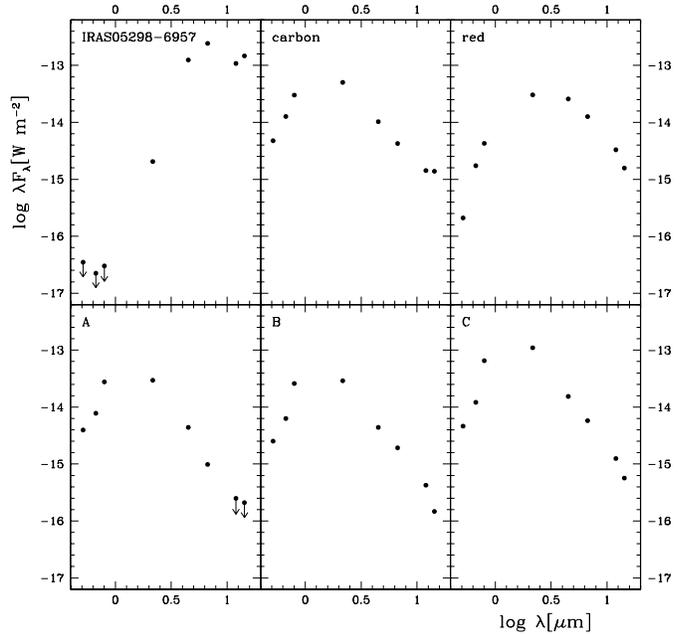,width=88mm}}
\caption[]{SEDs of the red stars near HS 327.}
\end{figure}

The photometric SEDs of the six red stars near HS 327 are plotted in Fig.\ 8.
IRAS05298$-$6957 is optically invisible because of the severe circumstellar
extinction by its massive dust envelope, whose emission produces a strong IR
excess. The cluster carbon star does not suffer from much circumstellar
extinction, but some IR excess emission is visible at the longest wavelengths.
The field carbon star (``red'') does experience significant extinction, but it
is yet unclear whether it is of interstellar or of circumstellar origin. The
three M giants are not noticeably reddened nor do they exhibit IR excess
emission, and their mass-loss rates must therefore be very low:
$\dot{M}\lsim10^{-8}$ M$_\odot$ yr$^{-1}$ (see van Loon et al.\ 1999b). The
derivation of accurate mass-loss rates requires additional photometry at
wavelengths of $\lambda\sim20$ to 60 $\mu$m.

\subsection{Luminosities}

%
%
\begin{table}
\caption[]{Spectral types and luminosities.}
\begin{tabular}{llrr}
\hline\hline
Star             &
Spectrum         &
log(L/L$_\odot$) &
M$_{\rm bol}$    \\
\hline
\multicolumn{4}{l}{\it Cluster members:} \\
IRAS05298$-$6957 &
OH/IR            &
$>4.3$           &
$<-6.0$          \\
carbon           &
C4.5             &
3.8              &
$-4.7$           \\
A                &
M3               &
3.7              &
$-4.5$           \\
\multicolumn{4}{l}{\it Field stars:} \\
red              &
C                &
3.6              &
$-4.3$           \\
B                &
M5               &
3.7              &
$-4.4$           \\
C                &
M7               &
4.1              &
$-5.6$           \\
\hline
\end{tabular}
\end{table}

Bolometric luminosities (Table 2) are estimated by integrating under the SED.
The luminosity derived for IRAS05298$-$6957 here is consistent with the more
complete derivation of the bolometric luminosity given by van Loon et al.\
(1999b): $M_{\rm bol}=-5.21$ to $-6.72$ mag. It is certainly the most luminous
AGB star in the field. Both the cluster and field carbon stars are a few times
less luminous, with luminosities typical for optically bright carbon stars
(Costa \& Frogel 1996). The red field star is less luminous than obscured
carbon stars detected by IRAS (van Loon et al.\ 1997, 1998, 1999a,b).

\subsection{Search for short-period variable stars}

The imaging data comprises six nights of intensive observation, which would,
in principle, allow the detection of variability on timescales of a few days
and with amplitudes of the order of several tenths of a magnitude, e.g.\
Cepheids. To this aim we investigated the 46 individual r-band images, using
only a single pass through DAOPHOT \& ALLSTAR. The $4^{\rm th}$ epoch
(December 25) was taken as a reference, for it contained the largest number of
extracted stars (4868). Of these, 3706 were recovered at least once more
during the time series. The relative photometric calibration between the
different epochs was improved by subtracting a baseline constructed from 146
stars that had been recovered at $\geq30$ epochs. All lightcurves were then
checked by eye for variability. No convincing variable star candidates could
be identified: a selection of $\sim50$ best cases were all consistent with
variations due to inaccuracies in the source extraction and photometry.

\section{Discussion}

\subsection{The age of the HS 327 cluster}

It is difficult to assess ages of the individual cluster stars. The locations
in the CMD of the cluster carbon star and star ``A'' suggest ages of $t<1$ Gyr
for these stars, depending on the exact amount of interstellar reddening. For
IRAS05298$-$6957 an age cannot be derived directly from the CMD because of the
difficulty to correct for its circumstellar extinction. It's luminosity of
$log(L/L_\odot)\gsim4.3$ and likely evolutionary state at the very tip of its
AGB suggests an age of $t\lsim500$ Myr (Bertelli et al.\ 1994).

The best indicators for the cluster age in the optical CMD are: (i) The
compact red clump at $(g-i)=1.1$, $i=18.2$ and $(i-K_{\rm s})=1.2$, $K_{\rm
s}=16.8$ represents stars with ages $1{\lsim}t<10$ Gyr; (ii) The bulk of the
Main Sequence stars for which photometry could be extracted have ages of a few
Gyr, but the data is incomplete for Main Sequence stars that are older; (iii)
An extension of the Main Sequence to younger stars with ages
$t{\gsim}2\times10^8$ yr; (iv) Luminous AGB stars with ages of a few
$\times10^8$ yr.

%
%
\begin{figure}[tb]
\centerline{\psfig{figure=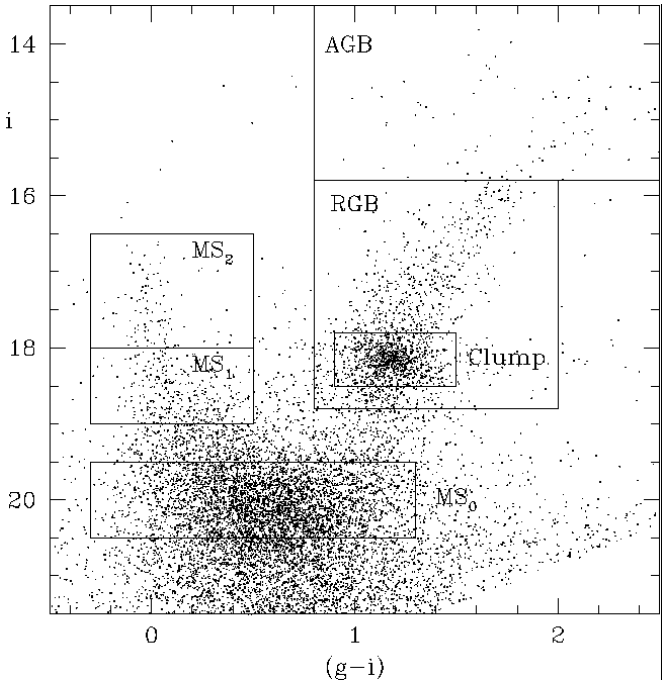,width=88mm}}
\caption[]{HRD with all stars in and around HS 327, with indicated the areas
for counting the different classes of stars.}
\end{figure}

%
%
\begin{table}
\caption[]{Relative number of the different classes of stars compared to their
total, $n=N_{\rm i}/N_{\rm tot}$, both for the cluster ($N_{\rm tot}=480$) and
the field ($N_{\rm tot}=6739$) around HS 327. The cluster-over-field density,
$\rho=N_{\rm i, cluster}/N_{\rm i, field}$, is corrected for the relative area
(0.000334 and 0.003401 $\sq^\circ$, respectively).}
\begin{tabular}{llll}
\hline\hline
                &
Cluster         &
Field           &
$\rho$          \\
\hline
MS$_0$          &
$0.529\pm0.041$ &
$0.541\pm0.011$ &
$0.71\pm0.05$   \\
MS$_1$          &
$0.056\pm0.011$ &
$0.049\pm0.003$ &
$0.84\pm0.17$   \\
MS$_2$          &
$0.025\pm0.007$ &
$0.014\pm0.001$ &
$1.33\pm0.41$   \\
Clump           &
$0.125\pm0.017$ &
$0.138\pm0.005$ &
$0.66\pm0.09$   \\
RGB             &
$0.233\pm0.024$ &
$0.247\pm0.007$ &
$0.68\pm0.07$   \\
AGB             &
$0.031\pm0.008$ &
$0.012\pm0.001$ &
$1.86\pm0.52$   \\
\hline
\end{tabular}
\end{table}

It is difficult to disentangle field and cluster due to the low contrast in
stellar density (Figs.\ 1 \& 6). In order to estimate whether the cluster
belongs to the dominant field population of a few Gyr, or to the sparse
younger field population of a few hundred Myr, statistics are obtained about
the relative frequency of stars of different classes as defined in the CMD
(Fig.\ 9): Main Sequence stars of a few Gyr (MS$_0$), $t\sim500$ Myr to 1 Gyr
(MS$_1$) and $t\sim200$ to 500 Myr (MS$_2$); red clump stars, RGB stars of
similar ages as the red clump stars, and AGB stars that generally represent
ages of several hundred Myr. The counts are normalised to their sum (Table 3),
and compared between cluster and field taking into account the differences in
area.

Although the statistics are not striking, the different classes of stars all
suggest that the cluster is associated with the younger population of a few
hundred Myr (that is also present in the field), and that the older population
of a few Gyr belongs entirely to the field. The statistics of the AGB stars
and youngest Main Sequence stars in particular favours the interpretation of
HS 327 having an age of $t\sim200$ Myr. This agrees with the age of
$t\sim160\pm20$ Myr as derived by Pietrzy\'{n}ski \& Udalski (2000a,b), which
happens to correspond to an epoch of intensified star formation that also
seems to have occurred in the field surrounding HS 327.

AGB stars seem to be relatively abundant within the cluster. As a simple
consistency check, the number $N_{\rm MS}$ of Main-Sequence stars within a
certain mass interval $[M_1,M_2]$ may be compared to the number $N_{\rm post}$
of post-Main Sequence stars within the mass interval $[M_3,M_4]$, where
$M_2=M_3=3.3$ M$_\odot$ the current (200 Myr) Main-Sequence Turn-Off mass, and
$M_4=4.0$ M$_\odot$ the Main-Sequence progenitor mass corresponding to the
current tip of the AGB (Bertelli et al.\ 1994). Assuming a Salpeter Initial
Mass Function (Salpeter 1955)
\begin{equation}
\frac{{\rm d}N}{{\rm d}M} = \xi M^{-2.35}
\end{equation}
the observed $N_{\rm MS}=12$ for $M_1=2.2$ M$_\odot$ (500 Myr: MS$_2$)
predicts $N_{\rm post}=5$, and the observed $N_{\rm MS}=81$ for $M_1=1.75$
M$_\odot$ (1 Gyr: MS$_1$+MS$_2$) predicts $N_{\rm post}=14$. These very rough
estimates agree reasonably well with the observed $N_{\rm post}\sim15$
(``AGB''). There are two TP-AGB stars (the carbon star and the OH/IR star)
where less than one would be expected, but this is a bias as the cluster was
studied exactly for reason of the presence of these two stars.

\subsection{Cluster binarity}

The LMC contains several populous binary clusters (Dieball et al.\ 2000;
Dieball \& Grebel 2000). HS 327 appears double (see also Pietrzy\'{n}ski \&
Udalski (2000a,b), showing that binarity also occurs amongst sparse open
clusters. The moderate age of HS 327 implies that such clusters can resist
tidal disruption for some while, although we do not know of course how many of
HS 327's members have already been stripped off in the past.

There is some indication for a difference in the stellar content of the West
and East components: HS 327-E is redder, or rather brighter in the $K_{\rm s}$
band. An analysis as performed in Table 3 but now comparing the East and West
components of HS 327 suggests that HS 327-W might be somewhat younger than HS
327-E. Perhaps star formation in the West part was triggered by momentum input
into the ISM by massive evolved stars in the East part. This result is not
statistically significant, though, and Pietrzy\'{n}ski \& Udalski (2000a,b)
derive ages that are equal to within $\sim50$ Myr.

\subsection{Progenitor masses of carbon stars and OH/IR stars}

The obscured oxygen-rich AGB star IRAS05298$-$6957 is more luminous than the
cluster carbon star, which in its turn is more luminous than the oxygen-rich
cluster star ``A''. Because these three stars have formed simultaneously, the
sequence in luminosities strongly suggests a sequence in evolutionary state as
a result of a sequence in progenitor mass, with IRAS05298$-$6957 the most
evolved and with the most massive progenitor. How can the sequence in
chemistry be understood?

Intermediate-mass AGB stars become carbon stars as $3^{\rm rd}$ dredge-up
becomes effective on the upper AGB. This process is counteracted upon in
massive AGB stars by carbon burning at the base of the convective mantle
(HBB). Clearly, the progenitor mass of the cluster carbon star must be high
enough to support $3^{\rm rd}$ dredge-up but too low for the onset of HBB,
whilst the progenitor of IRAS05298$-$6957 was sufficiently massive for HBB to
occur. Star ``A'' has yet to experience (enough) thermal pulses to become a
carbon star.

The cluster age of $t\sim200$ Myr implies a progenitor mass of $M_{\rm
MS}{\sim}4.0$ M$_\odot$ for stars currently at the tip of the AGB (Bertelli et
al.\ 1994). The range in progenitor masses of stars along the AGB is less than
0.1 M$_\odot$. Hence the first direct measurement for the threshold mass above
which HBB occurs indicates $M_{\rm HBB}{\gsim}4.0$ M$_\odot$.

This agrees well with current stellar evolution models. The lower and higher
mass limits for carbon star formation in the LMC are currently believed to be
$M_{\rm low}\sim1.2$ and $M_{\rm high}\sim4$ M$_\odot$, respectively
(Groenewegen \& de Jong 1993; Marigo et al.\ 1999). Ventura et al.\ (1999)
show that stars with $M_{\rm MS}{\sim}3.8$ M$_\odot$ first go through a short
(${\Delta}t\sim30,000$ yr) J-type ($^{13}$C-enhanced) carbon star phase before
HBB converts them back into oxygen-rich stars. Because of the short lifetime
and the lack of evidence for $^{13}$C enhancement, it is unlikely that the
cluster carbon star is currently going through this peculiar evolutionary
phase. It is possible that IRAS05298$-$6957 may have done so in the past,
though.

However, as the referee rightfully pointed out, it is difficult to reconcile
the low luminosity of the cluster carbon star with a 4 M$_\odot$ TP-AGB star.
Stellar evolution models (Marigo et al.\ 1999) predict that for a star with
$M_{\rm MS}{\sim}4.0$ M$_\odot$ the onset of the TP-AGB occurs around $M_{\rm
bol}\sim-6$ mag. The luminosity of the cluster carbon star of $M_{\rm
bol}\sim-4.7$ mag would rather suggest $M_{\rm MS}{\lsim}3.0$ M$_\odot$. The
star may have been fainter than average due to either stellar surface
pulsations, a post-TP luminosity dip or both, but it is unlikely that this
would account for a difference of $\sim1.3$ mag. It is also possible, but
unlikely, that the carbon star is not a cluster member. We therefore suggest
that current stellar models over-estimate the core mass (i.e.\ luminosity) at
which AGB stars with $M_{\rm MS}{\sim}4.0$ M$_\odot$ start to experience
$3^{\rm rd}$ dredge-up.

\begin{acknowledgements}
We are grateful for generous allocation by ESO of Director's
Discretionary Time. This work has benefitted from the use of the SIMBAD
database, operated at CDS, Strasbourg, France, and the Second Generation
Digitized Sky Survey, produced at the Space Telescope Science Institute under
U.S. Government grant NAG W-2166 and based on photographic data obtained using
the Oschin Schmidt Telescope on Palomar Mountain and the UK Schmidt Telescope.
We thank Fernando Comer\'{o}n for help with the SOFI observations and for
reading an earlier version of the manuscript. We also thank the referee for
her/his valuable suggestions. O Jacco se sente muito afortunado ser g\'{e}meos
unidos com o anjinho Joana.
\end{acknowledgements}

\appendix

\section{Photometric standard stars}

The standard stars used for the photometry in the Gunn system are listed in
Table A1, together with their magnitudes from the literature (Thuan \& Gunn
1976; Wade et al.\ 1979) and from our observations during the five nights from
December 25 to 29, 1996. Of these, HD 19445 and BD$+$21 607 are used as
primary standards because these are the only stars in our set that have known
i-band magnitudes. Ross 683 and especially Ross 889 showed the largest
discrepancies with respect to the full set of standard stars, and their use as
standard stars may therefore be questioned. Ross 374 and Ross 786 were found
to be in good agreement with their literature magnitudes. The magnitudes
determined here are accurate to $\sim0.01$ mag.

%
%
\begin{table}
\caption[]{Photometric standard stars for the Gunn system.}
\begin{tabular}{l|rrr|rrr}
\hline\hline
Star                            &
\multicolumn{3}{|c}{literature} &
\multicolumn{3}{|c}{observed}   \\
                                &
$g$                             &
$r$                             &
$i$                             &
$g$                             &
$r$                             &
$i$                             \\
\hline
BD$+$21 60\rlap{7}              &
 9.25                           &
 9.26                           &
 9.27                           &
 9.25                           &
 9.26                           &
 9.27                           \\
HD 19445                        &
 8.08\rlap{8}                   &
 8.07\rlap{0}                   &
 8.07\rlap{0}                   &
 8.09                           &
 8.07                           &
 8.07                           \\
Ross 374                        &
10.85\rlap{1}                   &
10.77\rlap{2}                   &
                                &
10.85                           &
10.77                           &
10.74                           \\
Ross 683                        &
11.40                           &
11.08                           &
                                &
11.38                           &
11.12                           &
11.04                           \\
Ross 786                        &
10.06                           &
 9.81                           &
                                &
10.03                           &
 9.81                           &
 9.73                           \\
Ross 889                        &
10.43                           &
10.51                           &
                                &
10.20                           &
10.27                           &
10.35                           \\
\hline
\end{tabular}
\end{table}


\begin{thebibliography}{}
\bibitem[1985]{}
Aaronson M., Mould J., 1985, ApJ 288, 551
\bibitem[1996]{}
Barnbaum C., Stone R.P.S., Keenan P.C., 1996, ApJS 105, 419
\bibitem[1994]{}
Bertelli G., Bressan A., Chiosi C., Fagotto F., Nasi E., 1994, A\&AS 106, 275
\bibitem[1993]{}
Boothroyd A.I., Sackmann I.-J., Ahern S.C., 1993, ApJ 416, 762
\bibitem[1996]{}
Costa E., Frogel J.A., 1996, AJ 112, 2607
\bibitem[2000]{}
Dieball A., Grebel E.K., 2000, A\&A 358, 897
\bibitem[2000]{}
Dieball A., Grebel E.K., Theis C., 2000, A\&A 358, 144
\bibitem[1999]{}
Egan M.P., et al., 1999, in: Astrophysics with Infrared Surveys: A Prelude to
SIRTF, eds.\ M.D. Bicay, R.M. Cutri \& B.F. Madore. ASP Conf.Ser. 177, p404
\bibitem[1994]{}
Fluks M.A., Plez B., Th\'{e} P.S., et al., 1994, A\&AS 105, 311
\bibitem[1998]{}
Frost C.A., Cannon R.C., Lattanzio J.C., Wood P.R., Forestini M., 1998, A\&A
332, L17
\bibitem[1993]{}
Groenewegen M.A.T., de Jong T., 1993, A\&A 267, 410
\bibitem[1966]{}
Hodge P.W., Sexton J.A., 1966, AJ 71, 363
\bibitem[1981]{}
Iben I., 1981, ApJ 246, 278
\bibitem[1983]{}
Iben I., Renzini A., 1983, ARA\&A 21, 271
\bibitem[1988]{}
Kontizas E., Metaxa M., Kontizas M., 1988, AJ 96, 1625
\bibitem[1993]{}
Kurucz R.L., 1993, Kurucz CD-ROM. Smithsonian Astrophysical Observatory,
Cambridge MA
\bibitem[1999]{}
Loup C., Cioni M.R., Blommaert J.A.D.L., 1999, in: The Universe as Seen by
ISO, eds.\ P. Cox \& M.F. Kessler. ESA-SP 427, p369
\bibitem[1998]{}
Marigo P., Bressan A., Chiosi C., 1998, A\&A 331, 564
\bibitem[1999]{}
Marigo P., Girardi L., Bressan A., 1999, A\&A 344, 123
\bibitem[1990]{}
Mathis J.S., ARA\&A 28, 37
\bibitem[1999]{}
Omont A., The ISOGAL Collaboration, 1999, in: The Universe as Seen by ISO,
eds.\ P. Cox \& M.F. Kessler. ESA-SP 427, p211
\bibitem[2000a]{}
Pietrzy\'{n}ski G., Udalski A., 2000a, Acta Astron.\ 50, 337
\bibitem[2000b]{}
Pietrzy\'{n}ski G., Udalski A., 2000b, Acta Astron.\ 50, 355
\bibitem[1995]{}
Price S.D., Witteborn F.C., 1995, in: Airborne Astronomy Symposium on the
Galactic Ecosystem: From Gas to Stars to Dust, eds.\ M.R. Haas, J.A. Davidson
\& E.F. Erickson. ASP Conf.Ser. 73, p685
\bibitem[1955]{}
Salpeter E.E., 1955, ApJ 121, 161
\bibitem[1998]{}
Schultheis M., Aringer B., H\"{o}fner S., J\/{o}rgensen U., 1998,in:
Asymptotic Giant Branch Stars, eds.\ T. Le Bertre, A. L\`{e}bre \& C.
Waelkens. ASP Conf.Ser., p614
\bibitem[1987]{}
Stetson P.B., 1987, PASP 99, 191
\bibitem[1997]{}
Tanab\'{e} T., Nishida S., Matsumoto S., et al., 1997, Nature 385, 509
\bibitem[1976]{}
Thuan T.X., Gunn J.E., 1976, PASP 88, 543
\bibitem[1999a]{}
Trams N.R., van Loon J.Th., Waters L.B.F.M., et al., 1999a, A\&A 344, L17
\bibitem[1999b]{}
Trams N.R., van Loon J.Th., Waters L.B.F.M., et al., 1999b, A\&A 346, 843
\bibitem[1985]{}
Turnshek D.E., Turnshek D.A., Craine E.R., Boeshaar P.C., 1985, An atlas of
digital spectra of cool stars. Western Research Company, Tucson
\bibitem[1997]{}
van Loon J.Th., Zijlstra A.A., Whitelock P.A., et al., 1997, A\&A 325, 585
\bibitem[1998]{}
van Loon J.Th., Zijlstra A.A., Whitelock P.A., et al., 1998, A\&A 329, 169
\bibitem[1999a]{}
van Loon J.Th., Zijlstra A.A., Groenewegen M.A.T., 1999a, A\&A 346, 805
\bibitem[1999b]{}
van Loon J.Th., Groenewegen M.A.T., de Koter A., et al., 1999b, A\&A 351, 559
\bibitem[1999]{}
Ventura P., D'Antona F., Mazzitelli I., 1999, ApJ 524, L111
\bibitem[1979]{}
Wade R.A., Hoessel J.G., Elias J.H., Huchra J.P., 1979, PASP 91, 35
\bibitem[1991]{}
Westerlund B.E., Azzopardi M., Breysacher J., Rebeirot E., 1991, A\&AS 91, 425
\bibitem[1983]{}
Wood P.R., Bessell M.S., Fox M.W., 1983, ApJ 272, 99
\bibitem[1992]{}
Wood P.R., Whiteoak J.B., Hughes S.M.G., et al., 1992, ApJ 397, 552
\end{thebibliography}
\end{document}